\documentclass[11pt,aps,onecolumn,eprint,prc,nofootinbib,noshowkeys,superscriptaddress,floatfix]{revtex4-1}
\usepackage{amsmath,amsfonts,amsthm,amssymb}
\usepackage[dvips]{graphics,graphicx}
\usepackage[usenames,dvipsnames]{color}
\definecolor{darkblue}{RGB}{0,0,196}
\definecolor{darkgreen}{RGB}{0,120,0}
\usepackage[colorlinks=true,linktocpage=true,linkcolor=darkblue,citecolor=red,urlcolor=darkblue]{hyperref}
\usepackage{cancel}
\usepackage{multirow}
\usepackage{longtable}
\usepackage{color}
\usepackage[normalem]{ulem}
\usepackage{hyperref}
\usepackage{bigints}
\usepackage{xparse}
\usepackage{physics}
\usepackage{verbatim}
\usepackage{minibox}
\usepackage{comment}
\usepackage{appendix}
\usepackage{mathtools}
\usepackage{slashed}
\usepackage{subfigure}

\newcommand{\bx}{\Bar{X}}
\newcommand{\bp}{\Bar{P}}
\newcommand{\bk}{\Bar{K}}
\newcommand{\bl}{\Bar{L}}
\newcommand{\bpsi}{\Bar{\psi}}
\newcommand{\tp}{\Tilde{p}_n}

\newcommand{\mc}{\mathcal{C}}

\newcommand{\sint}{~~{\mathclap{\textstyle\sum}\mathclap{\displaystyle\int}}}
\newcommand{\ab}[1]{\left\langle#1\right\rangle}

\begin{document}

\title{Energy-momentum correlators of fermions at finite temperature and density}

\author{Sourav Dey}
\email{sourav.dey@niser.ac.in}
\affiliation{School of Physical Sciences, National Institute of Science Education and Research, An OCC of Homi Bhabha Nuclear Institute, Jatni-752050, India.}

\author{Samapan Bhadury}
\email{samapan.bhadury@uj.edu.pl}
\affiliation{Institute of Theoretical Physics, Jagiellonian University, ul. St. \L ojasiewicza 11, 30-348 Krakow, Poland}

\author{Wojciech Florkowski}
\email{wojciech.florkowski@uj.edu.pl}
\affiliation{Institute of Theoretical Physics, Jagiellonian University, ul. St. \L ojasiewicza 11, 30-348 Krakow, Poland}

\author{Radoslaw Ryblewski}
\email{radoslaw.ryblewski@ifj.edu.pl}
\affiliation{Institute of Nuclear Physics, Polish Academy of Sciences, PL-31-342 Krakow, Poland}

\author{Amaresh Jaiswal}
\email{a.jaiswal@niser.ac.in}
\affiliation{School of Physical Sciences, National Institute of Science Education and Research, An OCC of Homi Bhabha Nuclear Institute, Jatni-752050, India.}
\affiliation{Institute of Theoretical Physics, Jagiellonian University, ul. St. \L ojasiewicza 11, 30-348 Krakow, Poland}

\begin{abstract}
Equal-time commutators of different components of the energy-momentum tensor at spatially separated points are calculated for a relativistic quantum Fermi gas at finite temperature and density. Different definitions of such components, also known as different pseudogauges, are used and smeared with a Gaussian profile characterized by the width $\sigma$. In this way, we introduce observables that may represent measurements of energy and momentum in a spatial region of size $\sigma$. We find that the obtained commutators are sensitive to the pseudogauge chosen if the probed systems or the spatial separation are small. The pseudogauge dependence is expected as different quantum operators are analyzed in this case. On the other hand, we find that for sufficiently large probed systems or with large separation, the studied commutators are pseudogauge independent. 
\end{abstract}

\maketitle

\section{Introduction}
\label{sec:Intro}

Recent measurements of spin polarization in relativistic heavy-ion collisions \cite{Liang:2004ph, Liang:2004xn, STAR:2017ckg, STAR:2018pps, STAR:2018fqv, Niida:2018hfw, STAR:2018gyt, ALICE:2019aid, STAR:2019erd, ALICE:2019onw, Singha:2020qns, Chen:2020pty, STAR:2020xbm, ALICE:2021pzu, STAR:2021beb, Mohanty:2021vbt, STAR:2022fan, STAR:2023eck} have sparked increased interest in the construction of a new framework of relativistic hydrodynamics that includes the effects of spin degrees of freedom. This framework, known as relativistic spin hydrodynamics, is currently being developed by several groups worldwide; for example, see Refs.~\cite{Voloshin:2004ha, Becattini:2007nd, Becattini:2009wh, Karabali:2014vla, Montenegro:2017rbu, Li:2017slc, Wang:2017jpl, Florkowski:2017ruc, Sun:2017xhx, Becattini:2017gcx, Florkowski:2017dyn, Florkowski:2018fap, Florkowski:2019qdp, Weickgenannt:2019dks, Kapusta:2019sad, Ayala:2019iin, Bhadury:2020puc, Ayala:2020ndx, Montenegro:2020paq, Weickgenannt:2020aaf, Speranza:2020ilk, Shi:2020htn, Bhadury:2020cop, Singh:2020rht, Hu:2021lnx, Bhadury:2021oat, Fu:2021pok, Becattini:2021suc, Becattini:2021iol, Hongo:2021ona, Singh:2021yba, Hu:2021pwh, Florkowski:2021wvk, Hongo:2022izs, Weickgenannt:2022zxs, Gallegos:2022jow, Bhadury:2022ulr, Weickgenannt:2022jes, Weickgenannt:2022jes, Weickgenannt:2022qvh, Sarwar:2022yzs, Biswas:2022bht, Biswas:2023qsw, Weickgenannt:2023nge, Hidaka:2023oze, Banerjee:2024xnd, Wagner:2024fhf, Lin:2024cxo, Yi:2024kwu, Tiwari:2024trl, Bhadury:2024ckc, Fang:2024sym}. In spin hydrodynamics, the inclusion of spin degrees of freedom is achieved by using the spin tensor as a new hydrodynamic current. However, this procedure introduces a problem related to the definitions of the energy-momentum and spin tensors: there exists an arbitrariness in their definitions, which leads to local redistribution of energy, linear momentum, and angular momentum. This arbitrariness (referred to as pseudogauge freedom) does not affect the total charges of conserved currents and the form of the conservation laws~\cite{Hehl:1976vr, Leader:2013jra}. However, the physical interpretation and consequences of pseudogauge dependence of hydrodynamic quantities are still not fully understood and are subject to intense investigations~\cite{Becattini:2011ev, Becattini:2011ev, Becattini:2012pp, Buzzegoli:2021wlg, Weickgenannt:2022jes, Dey:2023hft, Buzzegoli:2024mra}.
It is well known that for two physical quantities to be simultaneously observable, their operator representations must commute. The application of relativistic hydrodynamics to heavy-ion collisions is based on the assumption that the components of the energy-momentum tensor always commute. In this paper, we investigate the validity of this assumption and identify the conditions under which it is met. Moreover, the pseudogauge dependence of the commutator of energy-momentum tensor components is also analyzed. 

To analyze the problems outlined above, equal-time commutators (ETCs) of different components of the energy-momentum tensor at spatially separated points are calculated for a relativistic quantum Fermi gas at finite temperature and density. The method of the spectral functions in the imaginary time formalism is used~\cite{Laine:2016hma}. This allows us to identify areas where quantum effects play a significant role, pointing to the unreliability of classical hydrodynamic theories.

To determine a scale of coarse-graining over which the quantum effects become negligible, we consider subsystems of the full thermodynamic system~\cite{Coleman:2018mew}. We introduce a smooth Gaussian profile of the width $\sigma$ to define each subsystem, which leads to the smearing of the energy-momentum tensor components~\cite{Das:2020ddr, Das:2021aar}. Subsequently, we study the equal-time commutators of the smeared components of the energy-momentum tensor. 

We probe the properly scaled ETCs for several popular pseudogauge choices. In each of the considered cases, we study the transition of the system to the classical regime as a function of various scales of the theory such as mass, temperature, chemical potential, etc. We find that in the thermodynamic limit, where the smearing width becomes very large, $\sigma\rightarrow\infty$, all the ETCs vanish irrespective of the pseudogauge choice, which is in agreement with the previous studies \cite{Das:2020ddr, Das:2021aar}. Increasing the values of spatial separation, mass, or temperature, forces the system to tend towards the classical limit, again irrespective of the pseudogauge choice. On the other hand, the rate of approach to classical limit does depend on the specific pseudogauge choice. Finally, we find that the dependence of equal time commutators of the weighted energy-momentum tensor on chemical potential is not significant.

\textbf{Notation and conventions:} In this work, we use the following notation and conventions: Natural units $\hbar = c = k_B = 1$ are adopted and the metric tensor has the form $g^{\mu\nu} = \textrm{diag}(1, -1 , -1, -1)$. The Greek indices are reserved for four-vectors and run from 0 to 3, whereas Latin indices (like $j,k,l$) are used for three-vectors and run from 1 to 3. Furthermore, spinor indices are represented by the Latin indices from the beginning of the alphabet ($a,b,c,d,...$). Boldfaced letters denote three-vectors, $\textbf{x} = (x^1,x^2,x^3)$, whereas normal font letters denote four-vectors, $x = (x^0,x^1,x^2,x^3)$. We use $X^\mu \equiv \left( t, \textbf{x} \right)$ and $P^\mu \equiv \left( p^0, \textbf{p} \right)$ to denote the Minkowski position and momentum. For their Euclidean counterparts, we introduce two types of notation using the bar symbol. If the bar is put over the letter denoting the four-vector itself then we have $\Bar{X}^{\mu} \equiv \left( \tau, \textbf{x} \right)$ and $\Bar{P}^{\mu} \equiv \left( \Tilde{p}_n, \textbf{p} \right)$ for position and momentum, respectively~\footnote{We will often use $\tp = p_n + i \mu$ for the sake of simplicity, where $\mu$ and $p_n$ are the chemical potential and the Matsubara frequency of the Dirac fermions respectively.}. On the other hand, if the bar is put over the indices then we have $X^{\Bar{\mu}} \equiv \left( - i \tau, \textbf{x} \right)$ and $P^{\Bar{\mu}} \equiv \left( - i \Tilde{p}_n, \textbf{p} \right)$. For derivatives we use $\partial_\mu \equiv \left( \partial_t, \partial_{\textbf{x}} \right)$ for Minkowski coordinates and $\partial_{\Bar{\mu}} \equiv \left( i \partial_\tau, \partial_{\textbf{x}} \right)$ for Euclidean coordinates. The combined sum and integral notation is defined as $\sint_{P} \equiv T \sum_{p_n} \int_{\textbf{p}}$, where $\int_{\textbf{p}} \equiv \int \frac{d^3\textbf{p}}{\left(2\pi\right)^3}$. The dot symbol is used for three and four-vector dot products and should be understood from the context. Hence we have $A\cdot B = A_\mu B^\mu$ and $\textbf{a} \cdot \textbf{b} = a^i b^i$. Further, the hat symbol $\hat{A}$ denotes operators. We have also adopted Einstein's summation convention, where repeated indices are summed over.

\section{Thermal Commutators}
\label{sec:form}

For fields with finite spin, in addition to the energy-momentum tensor ($\hat{T}^{\mu\nu}$) and charge current ($\hat{N}^\mu$), we also have to consider the conservation of angular momentum ($\hat{J}^{\lambda,\mu\nu}$). The latter can be decomposed into orbital and spin parts as, $\hat{J}^{\lambda,\mu\nu} = \hat{L}^{\lambda,\mu\nu} + \hat{S}^{\lambda,\mu\nu}$, where the orbital part can be expressed in terms of the energy-momentum tensor as, $\hat{L}^{\lambda,\mu\nu} = x^\mu \hat{T}^{\lambda\nu} - x^\nu \hat{T}^{\lambda\mu}$. It has been known for a long time that there exists an ambiguity in the definitions of the energy-momentum tensor and spin current for systems with finite spin~\cite{Hehl:1976vr, Leader:2013jra}. This ambiguity is due to the invariance of equations of motion under the so-called pseudogauge transformations,
\begin{eqnarray}
\hat{T}^{\prime \mu \nu}&\!\!=\!\!& \hat{T}^{\mu\nu} + \frac{1}  {2} \partial_\lambda \left( 
\hat{\Phi}^{\lambda, \mu \nu} 
+\hat{\Phi}^{\nu, \mu \lambda}
+\hat{\Phi}^{\mu, \nu \lambda} \right), \nonumber \\
\hat{S}^{\prime \lambda, \mu \nu}&\!\!=\!\!&
\hat{S}^{\lambda, \mu \nu} - \hat{\Phi}^{\lambda, \mu \nu} + \partial_\rho \hat{Z}^{\mu\nu, \lambda \rho}. \label{eq:PG}
\end{eqnarray}
Here, $\hat{\Phi}^{\lambda, \mu \nu}$ and $\hat{Z}^{\mu\nu, \lambda \rho}$ are tensors known as superpotentials having the following symmetries,
\begin{eqnarray}
\hat{\Phi}^{\lambda, \mu \nu} &=& -\hat{\Phi}^{\lambda, \nu \mu}, \nonumber \\
\hat{Z}^{\mu\nu, \lambda \rho} &=& - \hat{Z}^{\nu\mu, \lambda \rho} \,\,\,=\,\,\, -\hat{Z}^{\mu\nu, \rho \lambda}. \label{eq:PGsymmetries}
\end{eqnarray}
In this work, we evaluate thermal correlators of the energy-momentum tensor for four different pseudogauge choices which are commonly used in the literature.
To determine the thermal correlators, we use the \emph{imaginary time formalism}~\cite{Laine:2016hma}. Subsequently, using these thermal correlators, we evaluate the thermal commutators of the energy-momentum tensor components in four pseudogauge choices - (i) Canonical, (ii) Belinfante, (iii) de Groot-van Leeuwen-van Weert (GLW) and, (iv) Hilgevoord-Wouthuysen (HW)~\cite{Speranza:2020ilk}. These commutators will allow us to understand which pseudogauge is more suitable for the application of classical effective theories such as relativistic hydrodynamics. We denote the Euclidean Dirac field operator by $\psi(\bx)$ and $\Bar{\psi}(\bx)$, which can be related to their counterparts in Fourier space as~\footnote{The definitions in Eqs.~\eqref{psi_E-FD} and \eqref{bpsi_E-FD} are in accordance to the Kubo-Martin-Schwinger (KMS) boundary conditions for fermions:
\begin{align*}
    \psi \left( - i \beta, \textbf{x}\right) = - e^{- \mu\beta} \psi \left(0, \textbf{x}\right)
    \quad{\rm and}\quad
    \bpsi \left( - i \beta, \textbf{x}\right) = - e^{\mu\beta} \bpsi \left(0, \textbf{x}\right),
\end{align*}
where $\beta = 1/T$ is the inverse temperature.
}\cite{Laine:2016hma,Kapusta:2023eix},
\begin{subequations}
\begin{align}
    \psi\left(\bx\right) &=\! \sint_{\{P\}}\! e^{i \bp \cdot \bx} \Tilde{\psi} \left(\bp\right) \!=\! \sint_{\{P\}} e^{i \tp \tau - i \textbf{p} \cdot \textbf{x}} \Tilde{\psi} \left(\bp\right), \label{psi_E-FD}\\
    \bpsi\left(\bx\right) &=\!\! \sint_{\{P\}}\!\! e^{-i \bp \cdot \bx} \Tilde{\Bar{\psi}} \left(\bp\right) \!=\! \sint_{\{P\}}\!\! e^{- i \tp \tau + i \textbf{p} \cdot \textbf{x}}\Tilde{\Bar{\psi}} \left(\bp\right), \label{bpsi_E-FD}
\end{align}
\end{subequations}
where the curly brackets in the subscript of sum integral notation imply the fermionic nature of the Matsubara frequencies. For an analogous sum integral over bosonic nature, we will not have the curly brackets. We will encounter four types of derivatives given below:
\begin{subequations}
    \begin{align}
        \partial_{\Bar{0}} \psi\left(\bx\right) &= \sint_{\{P\}} i \tp\, e^{i \bp \cdot \bx} \Tilde{\psi} \left(\bp\right) \label{db0-psi_E-FD}, \\
        \partial_{\Bar{0}} \Bar{\psi}\left(\bx\right) &= - \sint_{\{P\}} i \tp\, e^{- i \bp \cdot \bx} \Tilde{\Bar{\psi}} \left(\bp\right) \label{db0-bpsi_E-FD}, \\
        \partial_{j} \psi\left(\bx\right) &= - \sint_{\{P\}} i p_j\, e^{i \bp \cdot \bx} \Tilde{\psi} \left(\bp\right) \label{dj-psi_E-FD}, \\
        \partial_{j} \Bar{\psi}\left(\bx\right) &= \sint_{\{P\}} i p_j\, e^{- i \bp \cdot \bx} \Tilde{\Bar{\psi}} \left(\bp\right) \label{dj-bpsi_E-FD},
    \end{align}
\end{subequations}
where we have used the definitions, $\partial_{\Bar{0}} \equiv \frac{\partial}{\partial \tau}$ and, $\partial_{j} \equiv \frac{\partial}{\partial x^j}$.

Having defined all the notations associated with the fundamental fields, we will now start evaluating the thermal correlators of $T^{\mu\nu}$ for each of the pseudogauges from (i) to (iv).

\subsection{Canonical}
\label{ssec:for-Can}

The canonical energy-momentum tensor in Minkowski coordinates is given by \cite{Speranza:2020ilk},
\begin{align}
    \hat{T}^{C}_{\mu\nu} \left(X\right) &= \frac{i}{2} \bpsi\left(X\right)\, \gamma_\mu\, \overleftrightarrow{\partial_\nu} \psi \left(X\right), \label{T^C_mn-def-M}
\end{align}
where $\overleftrightarrow{\partial_\mu} = \overrightarrow{\partial_\mu} - \overleftarrow{\partial_\mu}$. In the Euclidean coordinates, this becomes
\begin{align}
    \hat{T}^{C}_{\mu\nu} \left(\bx\right) &= \frac{i}{2} \bpsi\left(\bx\right)\, \gamma_\mu\, \overleftrightarrow{\partial_{\Bar{\nu}}} \psi \left(\bx\right).
    \label{T^C_mn-def-E}
\end{align}
Therefore, the Euclidean correlation function of the energy-momentum tensor can be written as,
\begin{align}
    \ab{\hat{T}^{C}_{\mu\nu} \left(\bx\right) \hat{T}^C_{\alpha\beta} \left(\bx'\right)} 
    &= - \frac{1}{4} \left(\gamma_{\mu}\right)_{ab} \left(\gamma_{\beta}\right)_{cd} \overleftrightarrow{\partial_{\Bar{\nu}}}\, \overleftrightarrow{\partial_{\Bar{\alpha}'}}\ab{\bar{\psi}_{a}(\bx)\psi_{b}(\bx)\bar{\psi}_{c}(\bx')\psi_{d}(\bx')}\nonumber\\
    &=\! \frac{1}{4} \!\left(\gamma_{\mu}\right)_{ab} \!\left(\gamma_{\beta}\right)_{cd} \overleftrightarrow{\partial_{\Bar{\nu}}}\, \overleftrightarrow{\partial_{\Bar{\alpha}'}} \Big[G^{E}_{bc} \left(\bx',\bx\right) G^{E}_{da} \left(\bx,\bx'\right) \Big], \label{<TT>-C1}
\end{align}
where $\partial_{\Bar{\alpha}'} \equiv \partial/\partial X'^{\Bar{\alpha}}$. The angular brackets, $\ab{\left(\cdots\right)}$ imply the thermal vacuum expectation value of the quantity $\left(\cdots\right)$ at a constant temperature and chemical potential. The Euclidean Green's function is defined as,
\begin{align}
    G_{ab} \left(\bx',\bx\right) = \ab{\psi_a \left(\bx\right) \bpsi_{b} \left(\bx'\right)}. \label{G^E_ab-def}
\end{align}
The Euclidean Green's function can be computed for a given Lagrangian and expressed in terms of the spectral density function, $\rho_{ab}\left(\omega, \textbf{p}\right)$ as,
\begin{align}
    G_{ab} \left(\bx', \bx\right) = \sint_{\{\bp\}}e^{i\bp\cdot (\bx-\bx')} \int_{-\infty}^\infty \frac{d\omega}{\pi} \frac{\rho_{ab}\left(\omega, \textbf{p}\right)}{\omega + i \Tilde{p}_n}. \label{G-rho_rel}
\end{align}
In the case of free Dirac theory, the spectral density function is given by,
\begin{align}
    \rho_{ab}\left(\omega, \textbf{p}\right) &= \left(\slashed{P} + m\right)_{ab} \rho^0 \left(\omega, \textbf{p}\right), \label{rho_ab-def-Dirac}
    \end{align}
    with
\begin{align}
    \rho^0 \left(\omega, \textbf{p}\right) &= \left(\frac{\pi}{2 E_{\textbf{p}}}\right) \Big[ \delta \left(p^0 - E_{\textbf{p}}\right) - \delta \left(p^0 + E_{\textbf{p}}\right) \Big]. \label{rho^0-def-Dirac}
\end{align}
Using Eq.~\eqref{G-rho_rel}, we can rewrite Eq.~\eqref{<TT>-C1} as,
\begin{align}
    &\ab{\hat{T}^{C}_{\mu\nu} \left(\bx\right) \hat{T}^C_{\alpha\beta} \left(\bx'\right)} = \frac{1}{4} \sint_{\{\bp\bk\}} e^{i \left(\bp - \bk\right) \cdot \left(X - X'\right)} \nonumber\\
    &\quad\times \int_{-\infty}^\infty \frac{dp^0}{\pi} \int_{-\infty}^\infty \frac{dk^0}{\pi} \left(P^E_\nu + K^E_\nu\right) \left(P^E_\beta + K^E_\beta\right) \nonumber\\
    &\quad \times F^C_{\mu\alpha} \left(p^0,k^0, \textbf{p}, \textbf{k}, m\right) \frac{\rho^0\left(p_0, \textbf{p}\right) \rho^0\left(k_0, \textbf{k}\right)}{\left(p_0 + i \Tilde{p}_n\right) \left(k_0 + i \Tilde{k}_n\right)}, \label{<TT>-C2}
\end{align}
where $F^C_{\mu\alpha} \!\left(p^0,k^0, \textbf{p}, \textbf{k}, m\right) = \Tr \!\Big[ \gamma_\mu \!\left(\slashed{P} \!+\! m\right)\! \gamma_\alpha \!\left(\slashed{K} \!+\! m\right) \!\Big]$ $=\left(P_\mu K_\alpha + P_\alpha K_\mu\right) - g_{\mu\alpha} \left(P \cdot K - m^2 \right)$ and we have defined the quantity $P^E_\mu \equiv \left( - \Tilde{p}_n, i p_i\right)$, which can be viewed as the eigenvalue of the operator $\partial_{\Bar{\mu}}$ for the eigenfunction $e^{i \bp \cdot \bx}$, as can be seen from Eqs.~\eqref{db0-psi_E-FD}-\eqref{dj-bpsi_E-FD}. We can express this correlator from Eq.~\eqref{<TT>-C2} in terms of the momentum space 2-point Euclidean Green's function for the energy-momentum tensors as,
\begin{align}
    \ab{\hat{T}_{\mu\nu}\left(\bx\right) \hat{T}_{\alpha\beta} \left(\bx'\right)} &= \sint_{\bl} e^{i \bl \cdot \left(\bx - \bx'\right)} G_{\mu\nu,\alpha\beta} \left(\bl\right), \label{<TT>-G-E_C}
\end{align}
where $\bl^\mu = \bp^\mu - \bk^\mu$ has bosonic Matsubara frequencies i.e. $\bl^\mu = \left(l_n, \textbf{l}\right)$. Then we can write,
\begin{align}
    G_{\mu\nu,\alpha\beta}^C \left(\bl\right) &= \frac{1}{4} \sint_{\{\bp\}} \int_{-\infty}^\infty \frac{dp^0}{\pi} \int_{-\infty}^\infty \frac{dk^0}{\pi} \left(2 P^E_\nu - L^E_\nu\right) \nonumber\\
    &\quad\times \left(2 P^E_\beta - L^E_\beta\right) F^C_{\mu\alpha} \left(p^0,k^0, \textbf{p}, \textbf{p} - \textbf{l}, m\right) \nonumber\\
    &\quad\times\frac{\rho^0\left(p_0, \textbf{p}\right) \rho^0\left(k_0, \textbf{p} - \textbf{l}\right)}{\left(p_0 + i \Tilde{p}_n\right)\! \big[k_0 + i \left(\Tilde{p}_n - l_n\right)\!\big]}. \label{G^C(L)-def}
\end{align}
Using this momentum space 2-point Euclidean Green's function, we can obtain the corresponding spectral function as \cite{Laine:2016hma,Kovtun:2012rj},
\begin{align}
    &2i\, \rho_{\mu\nu,\alpha\beta} \left( \omega, \textbf{l} \right) = \Big[ G_{\mu\nu,\alpha\beta} \left( - i \omega + 0^+, \textbf{l}\right) \!-\! G_{\mu\nu,\alpha\beta} \left( - i \omega - 0^+, \textbf{l}\right) \!\Big]. \label{rho-G-rel}
\end{align}
Substituting Eq.~\eqref{G^C(L)-def} in Eq.~\eqref{rho-G-rel} and carrying out the thermal sums over the Matsubara frequency, $\Tilde{p}_n$ we can write,
\begin{widetext}
    \begin{align}
        \rho^{C}_{\mu\nu,\alpha\beta}(\omega, \textbf{l}) &= \pi \int_{\textbf{p}} \int_{-\infty}^\infty \frac{dp^{0}}{\pi} \int_{-\infty}^\infty \frac{dk^{0}}{\pi} \delta \left(k_0 - p_0 - \omega \right) \left(P_\nu^M + K_\nu^M \right) \left(P_\beta^M + K_\beta^M \right) \nonumber\\
        \times F^{C}_{\mu\alpha} &\left(p_0, k_0, \textbf{p}, \textbf{p} - \textbf{l}, m \right) \Big[ n_F (k_0 - \mu) - n_F (p_0 - \mu) \Big] \rho^0 (p_{0}, \textbf{p})\rho^0(k_0, \textbf{p} -\textbf{l}) + \big[ T = 0 \big], \label{rho^C-fin}
    \end{align}    \label{rhoom}
\end{widetext}
where we have defined $P^M_\mu = - i \left( p_0, - p_i\right)$ and $\big[ T = 0 \big]$ stands for the vacuum contribution. In the following, we have ignored all such vacuum contributions for the calculation of the thermal commutators. To compute the thermal sums we have used a general form of the Matsubara sum,
\begin{widetext}
    \begin{align}
        T\sum_{\{p_n\}} &\frac{ \prod_{j=1}^N \left(A\, \Tilde{p}_n - B\, l_n\right)^{q_j}}{\left(\omega_1 + i \Tilde{p}_n \right)} \frac{1}{\Big[\omega_2 + i \left(\Tilde{p}_n - l_n \right)\Big]} \label{General-TS}\\
        &= i^{\mathcal{N}} \left[ \frac{n_F\left(\omega_1 - \mu\right) - n_F\left(\omega_2 - \mu\right)}{\left(\omega_1 - \omega_2 + i l_n\right)} \right] \prod_{j=1}^N \Big[ \left(A \!-\! B\right) \omega_1 + B \omega_2 \Big]^{q_j}, \nonumber
    \end{align}
    where $\mathcal{N} = \sum_{j=1}^N q_j$.
\end{widetext}
We can now determine the thermal commutator from the spectral function from the relation,
\begin{align}
    \ab{\!\Big[\hat{T}_{\mu\nu} \!\left(X\right)\!, \hat{T}_{\alpha\beta} \!\left(X'\right) \!\Big]\!} \!&=\! 2 \!\int_{K}\! e^{i K \cdot (X - X')} \rho_{\mu\nu,\alpha\beta} \!\left(K\right), \label{<[T,T]>-rho_rel}
\end{align}
where we have used the definition, $\int_K \!\left(\cdots\right) \!\equiv\! \int \frac{dk^0}{2 \pi} \!\int_{\textbf{k}} \!\left(\cdots\right)$. Therefore, using Eq.~\eqref{rho^C-fin} in Eq.~\eqref{<[T,T]>-rho_rel} we can obtain the thermal commutator for the canonical energy-momentum tensors.
%
\subsection{Belinfante}
\label{ssec:for-Bel}

The energy-momentum tensor under Belinfante pseudogauge in Minkowski coordinates can be expressed by \cite{Speranza:2020ilk},
\begin{align}
    \hat{T}^B_{\mu\nu}(X) = \frac{1}{2} \left( \hat{T}^C_{\mu\nu} + \hat{T}^C_{\nu\mu} \right) .\label{T^B_mn-def-M}
\end{align}
Consequently, we can immediately write down the 2-point spectral function of the energy-momentum tensor under Belinfante pseudogauge as,
\begin{align}
    \rho^{B}_{\mu\nu,\alpha\beta}(\omega,\textbf{l}) &= \frac{1}{4} \Big[ \rho^{C}_{\mu\nu,\alpha\beta}(\omega,\textbf{l}) + \rho^{C}_{\mu\nu,\beta\alpha}(\omega,\textbf{l})  
     + \rho^{C}_{\nu\mu,\alpha\beta}(\omega,\textbf{l}) + \rho^{C}_{\nu\mu,\beta\alpha}(\omega,\textbf{l}) \Big]. \label{rho^B-fin}
\end{align}
Hence, we can use Eq.~\eqref{rho^B-fin} in Eq.~\eqref{<[T,T]>-rho_rel} to obtain the thermal commutator for the Belinfante energy-momentum tensors in Minkowski spacetime.

\subsection{GLW}
\label{ssec:for-GLW}

The energy-momentum tensor under GLW pseudogauge in Minkowski coordinates takes the form~\cite{Speranza:2020ilk},
\begin{align}
    \hat{T}^G_{\mu\nu}(X) = - \frac{1}{4 m} \bpsi\left(X\right) \overleftrightarrow{\partial_\mu}\, \overleftrightarrow{\partial_\nu} \psi\left(X\right) . \label{T^G_mn-def-M}
\end{align}
Following the steps in Sec.~\ref{ssec:for-Can} we can use Eq.~\eqref{T^G_mn-def-M} to obtain the spectral function from 2-point energy-momentum correlation function in the GLW pseudogauge by carrying out the fermionic thermal sums. This is found to be,
\begin{widetext}
    \begin{align}
        \rho^{G}_{\mu\nu,\alpha\beta}(\omega,\textbf{l}) &= -\frac{\pi}{4m^{2}} \int_{\textbf{p}} \int_{-\infty}^\infty \frac{dp^{0}}{\pi} \int_{-\infty}^\infty \frac{dk^{0}}{\pi}\delta(k_{0}-p_{0}-\omega)\label{rho^G-fin}\\
        &\times (P^{M}+K^{M})_{\mu}(P^{M}+K^{M})_{\nu}(P^{M}+K^{M})_{\alpha}\left(P^M + K^M \right)_{\beta} \nonumber\\
        &\times F^{G} \left(p_{0},k_{0},\textbf{p}, \textbf{p} - \textbf{l}, m\right) \Big[n_F \left(k_0 - \mu \right) - n_F \left(p_0 - \mu\right) \Big] \rho^{0}(p_{0},\textbf{p})\rho^{0} \left(k_{0},\textbf{p} - \textbf{l} \right) + \big[ T = 0 \big], \nonumber
    \end{align}
\end{widetext}
where $F^{G} \left(p_{0},k_{0},\textbf{p}, \textbf{k}, m\right) = P \cdot K + m^2$. Hence, we can use Eq.~\eqref{rho^G-fin} in Eq.~\eqref{<[T,T]>-rho_rel} to obtain the thermal commutator for the GLW energy-momentum tensors in Minkowski spacetime.

\subsection{HW}
\label{ssec:for-HW}

The energy-momentum tensor under HW pseudogauge in Minkowski coordinates is given by~\cite{Weickgenannt:2022zhj},
\begin{align}
    \hat{T}^H_{\mu\nu} \left(X\right) &= \hat{T}^G_{\mu\nu} \left(X\right) + \frac{1}{4 m} \partial_\mu\, \partial_\nu \Big[ \bpsi\left(X\right)\psi\left(X\right) \Big]  - g_{\mu\nu}\, \mathcal{L}_{KG},  \label{T^H_mn-def-M}
\end{align}
where, 
\begin{align}
    \mathcal{L}_{KG} \!=\! \frac{1}{2m} \Big[\! \left(\partial_\mu \bpsi \!\left(X\right) \right) \! \left(\partial^\mu \psi \!\left(X\right) \right) \!-\! m^2 \bpsi \!\left(X\right) \! \psi \!\left(X\right) \!\Big], \label{L_KG-Driac}
\end{align}
is the Klein-Gordon Lagrangian for Dirac fermions. Following the steps in Sec.~\ref{ssec:for-Can} we can use Eq.~\eqref{T^H_mn-def-M} to obtain the spectral function from 2-point energy-momentum correlation function in the HW pseudogauge by carrying out the fermionic thermal sums. This is found to be,
\begin{widetext}
    \begin{align}
        \rho^{H}_{\mu\nu,\alpha\beta} \left(\omega,\textbf{l}\right) &= - \frac{\pi}{4m^2} \int_{\textbf{p}} \int_{-\infty}^\infty \frac{dp^{0}}{\pi} \!\int_{-\infty}^\infty \frac{dk^{0}}{\pi} \delta\left(k_0 - p_0 - \omega\right)\nonumber\\
        &\times \!\Big[2 \left(P^M_\mu K^M_\nu \!+\! P^M_\nu K^M_\mu \right) + g_{\mu\nu} g^{\alpha\beta} \left(P^M_{\alpha} - K^M_{\alpha} \right) \! \left(P^M_{\beta} - K^M_{\beta} \right) \!\Big]\nonumber\\
        &\times \Big[2 \left(P^M_\alpha K^M_\beta + P^M_\beta K^M_\alpha \right) + g_{\alpha\beta} g^{\gamma\lambda} \left(P^M_{\gamma} - K^M_{\gamma} \right)\left(P^M_{\lambda} - K^M_{\lambda} \right) \Big] F^{H}(p_{0},k_{0},\textbf{p}, \textbf{p} - \textbf{l}, m) \nonumber\\
        &\times \Big[ n_F \left(k_0 - \mu \right) - n_F \left(p_0 - \mu\right) \Big] \rho^{0}(p_{0},\textbf{p})\rho^{0}(k_0, \textbf{p} - \textbf{l}) + \big[ T = 0 \big], \label{rho^H-fin}
    \end{align}
\end{widetext}
where $F^{H} \left(p_{0},k_{0},\textbf{p}, \textbf{k}, m\right) = P \cdot K + m^2$. Hence, we can use Eq.~\eqref{rho^H-fin} in Eq.~\eqref{<[T,T]>-rho_rel} to obtain the thermal commutator for the HW energy-momentum tensors in Minkowski spacetime.

\section{Effect of Finite Smearing}
\label{sec:EFS}

In the previous section, we have found the 2-point spectral functions of energy-momentum tensors and we saw that, by use of Eq.~\eqref{<[T,T]>-rho_rel}, we can obtain the thermal commutators. As mentioned earlier, we are interested in evaluating the commutators of the weighted energy-momentum tensors, which are defined as
\begin{align}
    \hat{T}^W_{\mu\nu} = \int d\textbf{x} W(\textbf{x} - \textbf{y}) \hat{T}_{\mu\nu} (t, \textbf{x}). \label{weighted-T}
\end{align}
We can use this to find the commutators of the weighted energy-momentum tensors at equal time under different pseudogauges as
\begin{align}
    &\ab{\!\Big[\hat{T}^W_{\mu\nu}(t, \textbf{y}),\hat{T}^W_{\alpha\beta}(t, \textbf{y}')\Big]\!} = \int\! d \textbf{x} \!\int\! d \textbf{x}'\, W \left(\textbf{y} - \textbf{x}\right)  W \left(\textbf{x}' - \textbf{y}' \right) \ab{\!\Big[\hat{T}_{\mu\nu} (t, \textbf{x}), \hat{T}_{\alpha\beta}(t, \textbf{x}') \Big]}. 
    \label{<[T^W,T^W]>-rho_rel}
\end{align}
In the following, for $W \left(\textbf{y} - \textbf{x}\right)$  we will use a Gaussian weight,  $W \left(\textbf{y} - \textbf{x}\right) = N_\sigma\, e^{- \left(\textbf{y} - \textbf{x}\right)^2/2 \sigma^2}$, with $N_\sigma$ being the normalization constant. Using this Gaussian profile and integrating out the spatial variables we get,
\begin{align}
    &\ab{\!\Big[\hat{T}^W_{\mu\nu}(t, \textbf{y}),\hat{T}^W_{\alpha\beta}(t, \textbf{y}')\Big]\!} = \int\frac{dk_{0}}{\pi} \int_{\textbf{k}}\, e^{-i \textbf{k} \cdot ( \textbf{y} - \textbf{y}')}   e^{-\textbf{k}^2 \sigma^2} \rho_{\mu\nu,\alpha\beta}(k_{0},\textbf{k}). \label{<[T^W,T^W]>-rho_rel2}
\end{align}
Using the results from Appendix~\ref{app:Parity} we note that only two types of commutators can have a non-zero values i.e., (i) $\ab{\!\Big[\hat{T}^W_{00}(Y), \hat{T}^W_{0j}(Y')\Big]\!}_{\!t=t'}$ and, (ii) $\ab{\!\Big[\hat{T}^W_{0j}(Y), \hat{T}^W_{k\ell}(Y')\Big]\!}_{\!t=t'}$ where $j,k,\ell$ take values $1,2,3$. For our purpose, it will suffice to consider the first case. In the following, setting $\textbf{y}' = 0$ and ignoring the vacuum parts in Eq.~\eqref{<[T^W,T^W]>-rho_rel2} we list the results for different pseudogauges\footnote{For the expressions of the correlators in the case of no smearing see Appendix~\ref{app:no-smear}.} as,
\begin{widetext}
    \begin{subequations}
        \begin{align}
            {\rm Canonical~:}\quad
            &\ab{\!\Big[T^{W,C}_{00}(t,\textbf{y}), T^{W,C}_{0j}(t, 0)\Big]\!}  = \frac{i\xi_{j}e^{-\frac{\xi^{2}}{4}}}{16(\sqrt{\pi})^{3}}\frac{1}{\sigma^{4}}{\cal H}\label{H<[T^C_00,T^C_0j]>-w}\\
            {\rm Belinfante~:}\quad
            &\ab{\!\Big[T^{W,B}_{00}(t,\textbf{y}), T^{W,B}_{0j}(t, 0)\Big]\!} = 0, \label{H<[T^B_00,T^B_0j]>-w}\\
            {\rm GLW~:}\quad
            &\ab{\!\Big[T^{W,G}_{00}(t,\textbf{y}), T^{W,G}_{0j}(t, 0)\Big]\!} = \frac{i\xi_{j}e^{-\frac{\xi^{2}}{4}}}{16(\sqrt{\pi})^{3}}\frac{1}{\sigma^{4}}\, \Big[{\cal H} \label{H<[T^G_00,T^G_0j]>-w} \\
             \quad
            & ~~~~~~~~~~~~~~~~~~~~~~~~
            +\frac{1}{(2\sigma m)^{2}}\left(\frac{5}{2}-\frac{\xi^2}{4}\right)\Big( \,2{\cal E} -\frac{5}{3}\,{\cal P} \Big)\Big], \nonumber\\
            {\rm HW~:}\quad
            &\ab{\!\Big[T^{W,H}_{00}(t,\textbf{y}), T^{W,H}_{0j}(t, 0)\Big]\!} = \frac{i\xi_{j}e^{-\frac{\xi^{2}}{4}}}{16(\sqrt{\pi})^{3}}\frac{1}{\sigma^{4}}{\cal H} \,, \label{H<[T^H_00,T^H_0j]>-w}
        \end{align}
    \end{subequations}
\end{widetext}
where, $\xi_{j}=\textbf{y}_{j}/\sigma$. Moreover, ${\cal H}={\cal H}(T,\mu,m),\,{\cal E}={\cal E}(T,\mu,m)$ and ${\cal P}={\cal P}(T,\mu,m)$ denote the enthalpy density, energy density and pressure respectively, which are defined with the help of the 1-point function, $\ab{\hat{T}_{\mu\nu} \left(t, \textbf{y}\right)} = \ab{\hat{T}^W_{\mu\nu} \left(t, \textbf{y}\right)}$.
Using the expressions from Eqs.~\eqref{T^C_mn-def-M}, \eqref{T^B_mn-def-M}, \eqref{T^G_mn-def-M}, and \eqref{T^H_mn-def-M} we can calculate the 1-point function for the operators $\hat{T}_{00}$ and $\hat{T}_{ii}$ for all the pseudogauges and find them all to be equal, and given by,
\begin{widetext}
    \begin{subequations}
\begin{align}
    {\cal E}  \!=\! \ab{\hat{T}_{00} \left(t, \textbf{y}\right)} \!=\! 2 \int_{\mathbf{p}}\!  \left(\textbf{p}^2 + m^2\right)^{1/2} \Big[ n_F \left(\sqrt{\textbf{p}^2 + m^2} - \mu\right) + n_F \left(\sqrt{\textbf{p}^2 + m^2} + \mu\right)\Big], \label{<T_00>-result}
\end{align}
\begin{align}
   {\cal P}  =  \frac{2}{3}\, \int_{\mathbf{p}}  \textbf{p}^2\left(\textbf{p}^2 + m^2\right)^{-1/2} \Big[ n_F \left(\sqrt{\textbf{p}^2 + m^2} - \mu\right) + n_F \left(\sqrt{\textbf{p}^2 + m^2} + \mu\right)\Big]. \label{<T_ii>-result}
\end{align}
\end{subequations}
\end{widetext}
Hence, enthalpy density can be expressed as ${\cal H}(T,\mu,m)={\cal E}(T,\mu,m)+{\cal P}(T,\mu,m)$.

We can immediately observe from the expressions of the commutators in Eqs.~\eqref{H<[T^C_00,T^C_0j]>-w}-\eqref{H<[T^H_00,T^H_0j]>-w} that in the limit of vanishing separation, i.e. $\textbf{y} \to 0$, all the commutators vanish which is obvious as essentially we end up evaluating the commutator of the same operator. On the other hand, in the limit of $\sigma \to \infty$, the commutators also vanish \cite{coleman2019quantum}, which can be again understood from the fact that in the case of infinite smearing width, the two operators with a finite spatial separation are indistinguishable and is equivalent to the case of $\textbf{y} \to 0$ limit.

Another interesting observation is the identical results for canonical and HW pseudogauges.

\section{Results and Discussions}
\label{sec:Res}

We note that the commutators obtained in Eqs.~\eqref{H<[T^C_00,T^C_0j]>-w}--\eqref{H<[T^H_00,T^H_0j]>-w} can be factorized as $\frac{1}{\sigma^{4}}\times f_1(\xi, \sigma m)\times f_2({\cal E, P, H})$,  where $f_1(\xi, \sigma m)$ is a function of the dimensionless variables, $\xi = \abs{\textbf{y}}/\sigma$ and $\sigma m$, while $f_2({\cal E, P, H})$ is a function of the thermodynamic variables. To study the qualitative features of the commutators, it will be more convenient to construct a dimensionless quantity through their normalization. In the present case, we will use the product of the enthalpy density (${\cal H}(T,\mu,m)$) and $l_T^{-4}$ as the normalizing quantity. Here $l_{T}$, which is the de Broglie thermal wavelength, is a natural length scale for the thermal medium (see  Appendix \ref{app:C-coeff} for more discussion on the properties of $l_T$.). This specific choice for normalization correctly captures the qualitative features in both the ultra-relativistic $(m\ll T)$ and the non-relativistic limit $(m\gg T)$. Thus, in the following, we consider a dimensionless quantity 
\begin{align}
    \mc_{\mu\nu,\alpha\beta} \left(\textbf{y}\right) =\frac{ \ab{\!\Big[\hat{T}^{W}_{\mu\nu}(t,\textbf{y}), \hat{T}^{W}_{\alpha\beta}(t, 0)\Big]\!}}{{\ab{{\cal H}(T,\mu,m)}}/l_{T}^{4}}. \label{Ratio-C}
\end{align}
Besides the dimensionless variable $\xi = \abs{\textbf{y}}/\sigma$, it is very convenient to introduce other dimensionless quantities: $\lambda = \sigma T$, $z = m/T$, $\alpha = \mu/T$, and $\mathcal{V}(z) = l_{T} T$. Thus, using Eqs.~\eqref{H<[T^C_00,T^C_0j]>-w}--\eqref{H<[T^H_00,T^H_0j]>-w} in Eq.~\eqref{Ratio-C} we can write:
\begin{subequations}
    \begin{align}
        \mc^C_{00,0j} \!\left(\textbf{y}\right) &= \mc^H_{00,0j} \!\left(\textbf{y}\right)  = \frac{i\, e_j\, \xi\, e^{- \xi^2/4}}{(\sqrt{\pi)^{3}}\,16 }\Bigg(\frac{\mathcal{V}(z)}{\lambda}\Bigg)^{4}, \label{C^C/H} \\
        \mc^B_{00,0j} \left(\textbf{y}\right) &= 0, \label{C^B} \\
        \mc^G_{00,0j} \left(\textbf{y}\right) &= \frac{i\, e_j\, \xi\, e^{- \xi^2/4}}{(\sqrt{\pi)^{3}}\,16 }\Bigg(\frac{\mathcal{V}(z)}{\lambda}\Bigg)^{4}\Bigg[ 1\,+ \frac{1}{(2\lambda\, z)^2} \!\left(\frac{5}{2} - \frac{\xi^2}{4} \right)  \frac{ 2{\cal E} -\frac{5}{3}{\cal P}  }{{\cal H} } \!\Bigg], \label{C^G}
    \end{align}
\end{subequations}
where $e_j \equiv y_j/\abs{\textbf{y}}$. 
    \begin{figure*}[t]
	    \centering 
	    \includegraphics[width=0.32\textwidth]{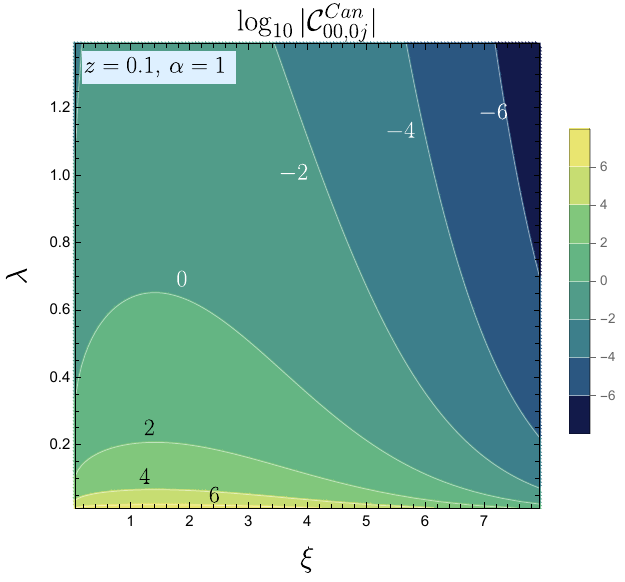} \hfill
        \includegraphics[width=0.32\textwidth]{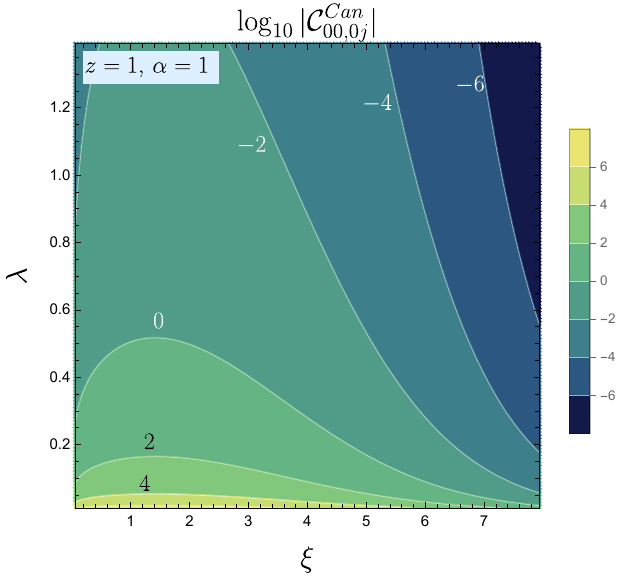} \hfill
        \includegraphics[width=0.32\textwidth]{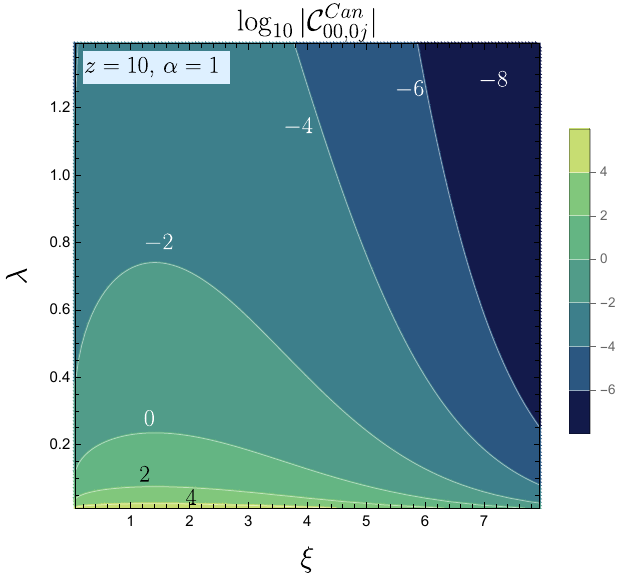} \\
	    \includegraphics[width=0.32\textwidth]{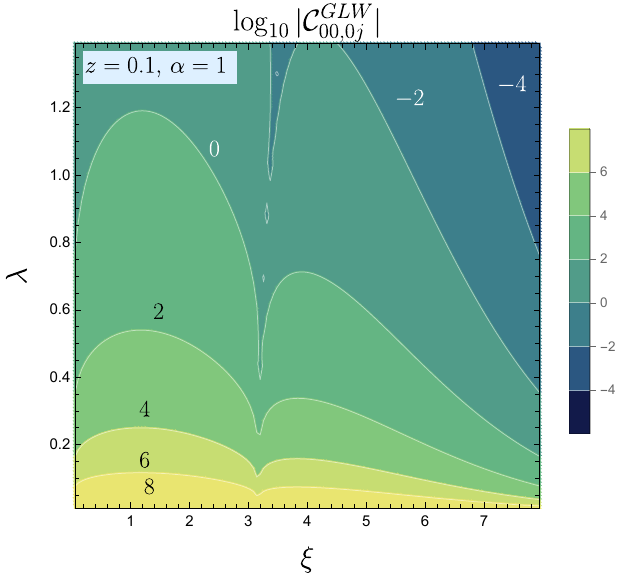} \hfill
        \includegraphics[width=0.32\textwidth]{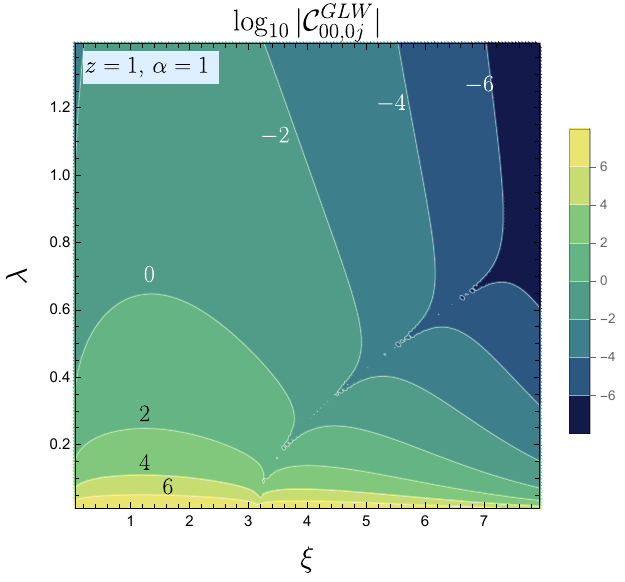} \hfill
        \includegraphics[width=0.32\textwidth]{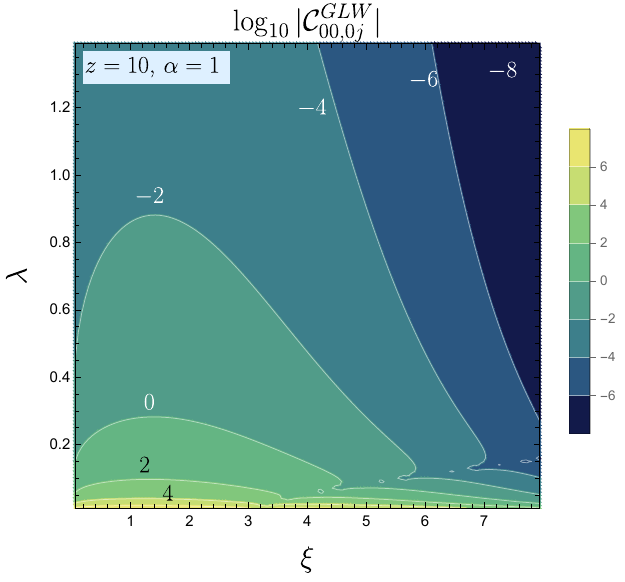}
	\caption{Contour plots of scaled commutator values of $\mc_{00,0j}$ from Eqs.~\eqref{C^C/H} and \eqref{C^G} as functions of the parameters $\xi$ and $\lambda$ for canonical and GLW pseudogauges respectively. Differently shaded regions denote the logarithmic value of $\mc_{00,0j}$. All the plots have been done for $\alpha = 1$ case. The first row represents the values for the canonical case and the second row represents the GLW case. From left to right, the $z=m/T$ value has been increased as  $z=0.1, 1, 10$. 
 }
	\label{fig_comp1}%
    \end{figure*}

\begin{figure}[t]
    \centering 
    \includegraphics[width=0.75\linewidth]{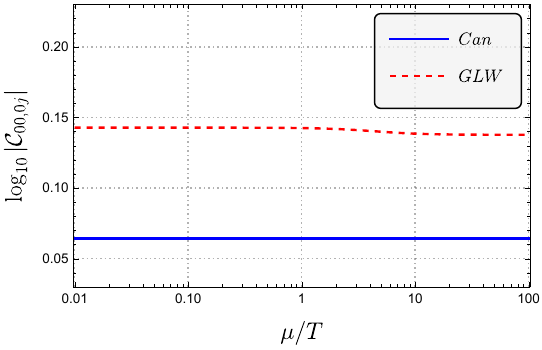}
    \caption{Variation of $\abs{\mc_{00,0j}}$ with $\mu/T$ for the Canonical and GLW cases at $\xi = 1$, $\lambda = 1$, and $z = 1$.}
    \label{fig_mu}%
\end{figure}

We note that among all possible components of  $\ab{\!\Big[T^W_{\mu\nu}(t, \textbf{y}),T^W_{\alpha\beta}(t, \textbf{0})\Big]\!}$, only $(00,0 i)$ and $(0 i,j k)$ give non-vanishing contributions and this can be shown by invoking the parity property of the commutators given by Eq.~(\ref{-O1O2}). It allows us to represent the ETC between two weighted energy-momentum tensors as an odd function of $\textbf{y}$, e.g.,  $(00,0 i)$ and $(0 i,j k)$ components can be expressed as $B\textbf{y}_{i}$ and $C\textbf{y}_{i} + D \textbf{y}_{i}\textbf{y}_{j}\textbf{y}_{k}$,  respectively, where the coefficients $B, C, D$ can be explicitly obtained as functions of $\xi$ and $\lambda$. Below we show our results for the $(00,0i)$ component of ETC for each pseudogauge choice only, since our results for the $(0i, jk)$ component are very similar~\footnote{ For the Belinfante choice, the $(00,0i)$ component vanishes, as given by Eq.~\eqref{C^B}, however, the results for the component $(0i, jk)$ do not vanish.}.

In Fig.~\ref{fig_comp1}, we plot the logarithm of the absolute value of the normalized correlation functions as given by Eqs.~\eqref{C^C/H} and \eqref{C^G}. Absolute values have been taken to avoid negative arguments of the logarithm. In the first row, we plot the normalized canonical correlation functions for different values of $z=m/T$ keeping the ratio $\alpha=\mu/T$ fixed, whereas, in the second row, we plot the same for the GLW pseudogauge. In both the canonical and GLW cases, a general trend is observed that the correlations become weaker as $z$ increases. 

We find the same trend of weakening correlation for the canonical case if $\lambda$ increases at fixed $\xi$. This behavior can be explained by noting that increasing $\lambda$ is equivalent to increasing temperature, $T$, or width, $\sigma$. It is expected that as the temperature increases, the system behaves classically with a vanishing correlation. The same holds for increasing the smearing width as for larger values of $\sigma$, the separation distance between the two spatial points where the operators are evaluated becomes effectively smaller compared to the $\sigma$ scale. In the case of the GLW, the dependence of the correlation functions on $\lambda$ is qualitatively similar to the canonical case, although there are quantitative differences.

Interestingly, there is a non-monotonic dependence of the canonical correlation function on $\xi$. At both small and large values of $\xi$ the canonical correlation function becomes weaker (one can check that it achieves a maximum for $\xi = \sqrt{2}$ which can be inferred from Eq.~\eqref{C^C/H}). This is an expected behavior as for very small values of $\xi$, the spatial separation of the two operators is negligible and we essentially compute the commutator of two identical operators. On the other hand, the large $\xi$ values are equivalent to a large spatial separation, where it is natural to expect that the influence of one operator on another is minimal. The $\xi$ dependence of the GLW case is very non-trivial. In particular, one may notice that there are two distinct peaks at different values of $\xi=\xi_\pm$ given by the formula
\begin{align}
    \xi_\pm = \sqrt{\frac{2}{M}} \left[1+4 M\pm\sqrt{1+M(6+11M)}\right]^{1/2}, \label{GLW-xi_pm}
\end{align}
which is obtained from Eq.~\eqref{C^G} with $M= \Big(2{\cal E} -\frac{5}{3}{\cal P} \Big)/(4\lambda^2 z^2{\cal H})$.
We note that there is a saddle formed between the two peaks. It can be noted from Eq.~\eqref{C^G} that a zero line ($\xi_0$) may exist in the $\lambda-\sigma$ plane of the correlation function. The locus of this line in the $\lambda-\xi$ plane is given by the equation
\begin{align}
    \xi_0
    = \sqrt{5 + \frac{2}{M}}.
\end{align}
From Fig.~\ref{fig_comp1} one may see that the quantum effects cannot be ignored for $\sigma$ smaller than about $1$ fm for all pseudogauges. This highlights a clear limitation for applying relativistic hydrodynamics at scales smaller than 1 fm. On the other hand, it is important to note that in systems produced in heavy-ion collisions, higher temperatures correspond to the early stages of evolution, during which the system is believed to be far from equilibrium. In such conditions, the current imaginary time formalism may not be a reliable approach. Consequently, more sophisticated calculations that extend beyond the simplistic model of a relativistic Fermi gas are necessary to address this issue in greater depth.

While in Fig.~\ref{fig_comp1} different correlators are studied as functions of $z, \lambda$, and $\xi$ keeping $\alpha$ fixed, in Fig.~\ref{fig_mu} we plot the correlators as functions of $\alpha=\mu/T$. As there is no $\mu$ dependence in the canonical correlator, which can be seen from Eq.~\eqref{C^C/H}, we observe a horizontal line indicating the constant value of the canonical correlator at fixed values of $z, \lambda$, and $\xi$. On the other hand, the GLW correlator depends on $\alpha$ very weakly and decreases very slowly with increasing $\alpha$.

Finally, we also note that at the fixed value of $\alpha$, with the increasing mass the magnitude of the correlation becomes smaller at a slower rate in the $\xi-\lambda$ plane, see Fig.~\ref{fig_comp1}. This is because, in the large mass limit, the system's dynamics fall in the non-relativistic regime, where the thermal de Broglie wavelength is $l_{T}\sim\frac{1}{\sqrt{m T}}\sim\frac{1}{T\sqrt{z}}$. Consequently, a large mass corresponds to a system with a small de Broglie wavelength which in turn implies classical theories can describe the dynamics of the system at a smaller lengthscale\footnote{On the other hand, in the ultra-relativistic limit or a very small mass limit, the thermal wavelength scales as $l_{T}\sim\frac{1}{T}$.}. Hence, for a fixed temperature $T$, the magnitude of the correlation becomes significantly smaller at the larger values of the particle's mass.  In Fig.~\ref{fig_mass} one can see that the canonical and GLW correlation functions decrease fast in the non-relativistic regime. However, at small values of $m$, the behavior of the two correlators differs. For the canonical case, it remains finite, but for the GLW case it diverges due to the presence of a term proportional to $z^{-2}$ in Eq.~\eqref{C^G}.
\section{Conclusions and Outlook}
\label{sec:C&O}

In this work, we have calculated the equal-time commutators of two spatially separated energy-momentum tensors for a relativistic Fermi gas at finite temperature and density with different choices of pseudogauges. We used smeared operators, with a Gaussian profile characterized by the width $\sigma$, to introduce observables that may represent measurements of energy and momentum in a spatial region of size $\sigma$. Our findings can be summarized by the following points:

\begin{enumerate}

\item By involving the symmetry properties of the correlation functions under parity operation we find that the only non-vanishing terms are $[\hat{T}^{00},\hat{T}^{0i}]$ and $[\hat{T}^{0i},\hat{T}^{jk}]$, i.e., the tensor components with odd number of spatial indices.

\item The general behavior found for the considered pseudogauges is that the quantum effects play a less significant role if either the temperature or spatial separation is increased.

\item Our study shows that for very large values of $\lambda = \sigma \,T$, with fixed temperature $T$, the quantum effects wash out similarly for all considered pseudogauges. However, an interesting feature is that for small values of $\lambda$ and fixed $T$, the choice of pseudogauge does matter --- the quantum effects get suppressed at larger values of $\xi= \abs{\textbf{y}}/\sigma$ for the GLW case as compared to the HW and Canonical cases.  At $\lambda\sim 1$ the magnitude of the correlation function for canonical as well as HW pseudogauge are $100$ times smaller than for the GLW case for all values of $\xi$.

\item Irrespective of the pseudogauge used, at a small $z=m/T$ ratio we find that the quantum effects cannot be ignored for $\sigma \sim 1$~fm. This may indicate potential problems for using relativistic hydrodynamics \cite{Spalinski:2016fnj, Ambrus:2022qya} at the scales of about 1 fm and smaller, although, more realistic calculations are needed in this case \cite{Romatschke:2023ztk}, which depart from a simple picture of a relativistic Fermi gas. 

\item We also note that at the fixed value of $\alpha$, in the large mass limit, the system becomes non-relativistic and the thermal de Broglie wavelength depends on mass as well as temperature. On the other hand, in the ultra-relativistic limit, the mass scale loses its relevance, and the thermal wavelength depends only on the temperature of the system. This interesting property of the thermal de Broglie wavelength shows the motivation behind the particular choice of normalization in Eq.~\eqref{Ratio-C}.

\end{enumerate}

\section{Acknowledgment}
\label{sec:ack}

S.D. acknowledges financial support from an academic cooperation agreement between the National Institute of Science Education and Research, Jatni and Jagiellonian University, Krakow. A.J. acknowledges the kind hospitality of IIT Gandhinagar where part of this work was carried out. This work was supported in part by the Polish National Science Centre Grants No. 2022/47/B/ST2/01372 (W.F.) and No. 2018/30/E/ST2/00432 (R.R., S.D. and A.J.).

\appendix
\section{\texorpdfstring{$\sigma \to 0$}{~} Case:}
\label{app:no-smear}

The expressions of the thermal commutators of the energy-momentum tensor for different pseudogauges in absence of any smearing are given by,
\begin{figure}[t]
    \centering 
    \includegraphics[width=\linewidth]{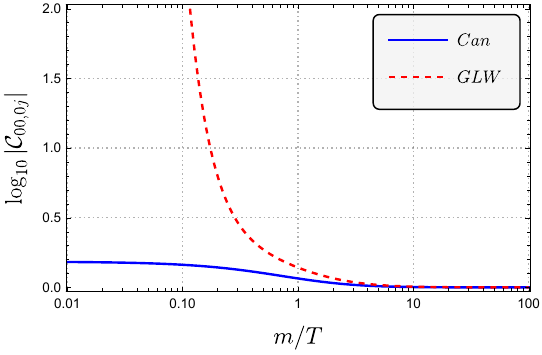}
    \caption{Variation of $\log_{10}\abs{\mc_{00,0j}}$ Canonical and GLW cases with $m/T$ at the values, $\xi = 1$, $\lambda = 1$ and, $\alpha = 1$.}
    \label{fig_mass}%
\end{figure}
\begin{widetext}
    \begin{subequations}
        \begin{align}
            {\rm Canonical~:}\quad
            &\ab{\!\Big[\hat{T}^{C}_{00}(t,\textbf{y}), \hat{T}^{C}_{0j}(t, 0)\Big]\!} = - \ab{\!\Big[\hat{T}^{C}_{00}(Y), \hat{T}^{C}_{j0}(0)\Big]\!} = i \partial_{\textbf{y}^{j}}\big(\delta^{3}(\textbf{y})\big)\mathcal{H} , \label{<[T^C_00,T^C_0j]>0-w}\\
            {\rm Belinfante~:}\quad
            &\ab{\!\Big[\hat{T}^{B}_{00}(t,\textbf{y}), \hat{T}^{B}_{0j}(t, 0)\Big]\!} = 0, \label{<[T^B_00,T^B_0j]>0-w}\\
            {\rm GLW~:}\quad
            &\ab{\!\Big[\hat{T}^{G}_{00}(t,\textbf{y}), \hat{T}^{G}_{0j}(t, 0)\Big]\!} = i\,\partial_{\textbf{y}^{j}}\big(\delta^{3}(\textbf{y})\big)\mathcal{H}  \label{<[T^G_00,T^G_0j]>0-w}   +\frac{i}{ 4m^{2}}\partial_{\textbf{y}^{j}}\partial^{2}_{\textbf{y}}\big(\delta^{3}(\textbf{y})\big)\Big(2\mathcal{E} -\frac{5}{3}\mathcal{P} \Big),   \\
            {\rm HW~:}\quad
            &\ab{\!\Big[\hat{T}^{H}_{00}(t,\textbf{y}), \hat{T}^{H}_{0j}(t, 0)\Big]\!} = i \partial_{\textbf{y}^{j}}\big(\delta^{3}(\textbf{y})\big)\mathcal{H}  , \label{<[T^H_00,T^H_0j]>0-w}
        \end{align}
    \end{subequations}
\end{widetext}
\section{Vanishing of commutators from parity}
\label{app:Parity}

In this section, we show that the condition for having a non-zero equal time commutator of a set of arbitrary bosonic-hermitian operator $\hat{\mathcal{O}}_{1}(t,\textbf{y})$ and $\hat{\mathcal{O}}_{2}(t,\textbf{y}')$ is that both the operators must have opposite parity structure. Let us denote the commutator of two such bosonic-hermitian operators, $\hat{\mathcal{O}}_{1}(t,\textbf{y})$ and $\hat{\mathcal{O}}_{2}(t,\textbf{y}')$ by \cite{Kovtun:2012rj},
\begin{align}
    \rho_{\mathcal{O}_{1}\mathcal{O}_{2}}(\textbf{y}-\textbf{y}')=\ab{\!\Big[\hat{\mathcal{O}}_{1}(t, \textbf{y}),\hat{\mathcal{O}}_{2}(t, \textbf{y}')\Big]\!}.\label{O1O2}
\end{align}
Let us consider the hermitian conjugate of Eq.~\eqref{O1O2} as,
\begin{align}
    \rho^\dag_{\mathcal{O}_{1}\mathcal{O}_{2}}(\textbf{y}-\textbf{y}') &= \ab{\!\Big[\hat{\mathcal{O}}_{1}(t, \textbf{y}),\hat{\mathcal{O}}_{2}(t, \textbf{y}')\Big]\!}^\dag \nonumber\\
    &= - \ab{\!\Big[\hat{\mathcal{O}}_{1}(t, \textbf{y}),\hat{\mathcal{O}}_{2}(t, \textbf{y}')\Big]\!} \nonumber\\
    &= - \rho_{\mathcal{O}_{1}\mathcal{O}_{2}}(\textbf{y}-\textbf{y}'). \label{O1O2*}
\end{align}
In the Fourier space this translates to,
\begin{align}
    \rho_{\mathcal{O}_{1}\mathcal{O}_{2}}(\omega,\textbf{k})=-\rho_{\mathcal{O}_{1}\mathcal{O}_{2}}(\omega,-\textbf{k}),\label{FO1O2}
\end{align}
If the underlying microscopic theory of the system remains invariant under parity transformation ($\mathcal{P}$),
then starting from Eq.~\eqref{O1O2} using Eq.~\eqref{FO1O2} we find,
\begin{align}
    \rho_{\mathcal{O}_{1}\mathcal{O}_{2}}(\textbf{y}-\textbf{y}')&=-\rho_{\mathcal{O}_{1}\mathcal{O}_{2}}(\textbf{y}'-\textbf{y}).\label{-O1O2}
\end{align}
Let us consider transformation of these operators under parity \cite{Sakurai:1994},
\begin{align}
    \mathcal{P} \hat{\mathcal{O}}_{a}(t,\textbf{x})P^{-1}=\eta_{a}\hat{\mathcal{O}}_{a}(t,-\textbf{x}), \label{PO_a}
\end{align}
where $a = 1,2$ in the present case and $\eta_a$ denotes the eigenvalue of the corresponding operator $\hat{\mathcal{O}}_a$. 

Therefore we can write,
\begin{align}
    \rho_{\mathcal{O}_{1}\mathcal{O}_{2}}(\textbf{y}-\textbf{y}') &= \ab{\!\Big[\hat{\mathcal{O}}_{1}(t, \textbf{y}),\hat{\mathcal{O}}_{2}(t, \textbf{y}')\Big]\!} \nonumber\\
    &= \ab{\!\Big[\hat{\mathcal{O}}_{1}(t, -\textbf{y}'),\hat{\mathcal{O}}_{2}(t, -\textbf{y})\Big]\!}\nonumber\\
    &=\eta_{1}\eta_{2}\ab{\!\Big[\hat{\mathcal{O}}_{1}(t, \textbf{y}'),\hat{\mathcal{O}}_{2}(t, \textbf{y})\Big]\!}\nonumber\\
    &=\eta_{1}\eta_{2}\rho_{\mathcal{O}_{1}\mathcal{O}_{2}}(\textbf{y}'-\textbf{y})\nonumber\\
    &= - \eta_{1}\eta_{2}\rho_{\mathcal{O}_{1}\mathcal{O}_{2}}(\textbf{y}-\textbf{y}'). \label{O1O2-parity}
\end{align}
To obtain the second equality we have used the property of translational invariance in spacetime, whereas the third equality is ensured by recalling the property of cyclic permutation under trace. The last step is obtained by using Eq.~\eqref{-O1O2}.

From Eq.~\eqref{O1O2-parity} we can conclude, that the commutator will vanish if both the operator has same parity eigenvalue and, only the commutators with opposite parity can have non-trivial values.

Similar conclusions can be drawn for the commutator of the Gaussian-weighted operators as,
\begin{align}
    &\ab{\!\Big[\hat{\mathcal{O}}_{1}^W(t, \textbf{y}),\hat{\mathcal{O}}_{2}^W(t, \textbf{y}')\Big]\!} = \int\! d \textbf{x} \!\int\! d \textbf{x}'\, W \left(\textbf{y} - \textbf{x}\right) \nonumber\\
    &\qquad\times W \left(\textbf{x}' - \textbf{y}' \right) \ab{\!\Big[\hat{\mathcal{O}}_{1} (t, \textbf{x}), \hat{\mathcal{O}}_{2}(t, \textbf{x}') \Big]}. 
    \label{<[O^W,O^W]>-rho_rel}
\end{align}
In our present work, we may thus conclude that the only commutators that can have a non-zero value are, (i) $\ab{\!\Big[\hat{T}^W_{00}(Y), \hat{T}^W_{0j}(Y')\Big]\!}_{\!t=t'}$ and, (ii) $\ab{\!\Big[\hat{T}^W_{0j}(Y), \hat{T}^W_{k\ell}(Y')\Big]\!}_{\!t=t'}$ where $j,k,\ell$ takes values $1,2,3$.

\section{Thermal length}\label{app:C-coeff}
A characteristic length scale can be constructed for a thermal system, which can qualitatively give a criterion for a reason whether this system is in a quantum or classical regime. If our considered subsystem's size $\sigma$ is comparable to the thermal wavelength, then the quantum effects become important. On the other hand, the thermal wavelength contains information on the average kinematic energy for a non-interacting theory of the quasi-particle distribution. From the phase-space definition, we can write it as,
\begin{align}
    \mathcal{Z}(T,m)=\int\frac{d^{3}\textbf{x}\,d^{3}\textbf{p}}{(2\pi)^{3}}e^{-\beta H_{\rm kin}}=\frac{V}{l^{3}_{T}}.\label{thwl}
\end{align}
Here $H_{\rm kin}$ is the kinetic energy of the particle, for a relativistic particle $H_{\rm kin}+m=\sqrt{\textbf{p}^{2}+m^{2}}$. Therefore, from the definition of thermal length Eq.~\eqref{thwl}, one can show that 
\begin{align}
    l_{T}^{3}=\frac{1}{T^{3}}\frac{2\pi^{2}e^{-z}}{z^{2}K_{0}(z)+2 z K_{1}(z)}.
\end{align}
This result can be checked in two limits, in the ultra-relativistic limit ($z\rightarrow0$), thermal wavelength goes as $l_{T}=\frac{\pi^{2/3}}{T}\sim \frac{1}{T}$. On the other hand, for non-relativistic limit ($z\rightarrow\infty$) this goes as $l_{T}=\sqrt{\frac{2\pi}{mT}}\sim\frac{1}{T\sqrt{z}}$. For the sake of convenience, we express thermal wavelength as,
\begin{align}
    l_{T}=\frac{1}{T}\Big(\frac{2\pi^{2}e^{-z}}{z^{2}K_{0}(z)+2 z K_{1}(z)}\Big)^{\frac{1}{3}}=\frac{\mathcal{V}(z)}{T}.
\end{align}
Here, $\mathcal{V}(z)$ shows asymptotics as $\sim z^{0}$ and $\sim z^{-\frac{1}{2}}$  for $z\rightarrow0$ and $z\rightarrow\infty$, respectively.
These asymptotics help to identify the behavior of the correlation function at the small and large values of $z$ consecutively, which can be shown,
\begin{align}
    \mc^C_{00,0j} \!\left(\textbf{y}\right) &= \mc^H_{00,0j} \!\left(\textbf{y}\right)\propto e_{j}\xi\,e^{-\frac{\xi^{2}}{4}}\lambda^{-4}\,,\\
    \mc^G_{00,0j} \left(\textbf{y}\right) &\propto e_j\, \xi\, e^{- \xi^2/4}\lambda^{-6}z^{-2}\!\left(\frac{5}{2} - \frac{\xi^2}{4} \right)\frac{\mathcal{P}}{\mathcal{H}}\,,
\end{align}
and 
\begin{align}
    \mc^C_{00,0j} \!\left(\textbf{y}\right) &= \mc^H_{00,0j} \!\left(\textbf{y}\right)\propto e_{j}\xi\,e^{-\frac{\xi^{2}}{4}}\lambda^{-4}z^{-2}\,,\\
    \mc^G_{00,0j} \left(\textbf{y}\right) &\propto e_j\, \xi\, e^{- \xi^2/4}\lambda^{-4}z^{-2}\!\Bigg[1+\left(\frac{5}{2} - \frac{\xi^2}{4} \right)\frac{\mathcal{E}}{2\lambda^{2}\,\mathcal{H}}\Bigg]\, .
\end{align}
\bibliography{ref}
\end{document}